\documentclass[manuscript=article]{achemso}
\usepackage[version=3]{mhchem}
\setkeys{acs}{articletitle=true}
\usepackage{graphicx, setspace}
\usepackage{hyperref,xcolor}
\usepackage{multicol}
\usepackage{textcomp,gensymb}
\usepackage[export]{adjustbox}
\usepackage{float,flushend,bm}
\usepackage[utf8]{inputenc}
\usepackage[T1]{fontenc}
\usepackage{mathptmx,balance}
\usepackage{afterpage}
\usepackage{tabularx,ragged2e,booktabs,caption}
\newcolumntype{C}[1]{>{\Centering}m{#1}}

\SectionNumbersOn

\author{Albin Joy}
\affiliation[IITT]
{Department of Chemistry, Indian Institute of Technology Tirupati, Yerpedu, Tirupati, Andhra Pradesh, India. 517619.}
\author{Anand Srivastava}
\affiliation[IISC]
{Molecular Biology Unit, Indian Institute of Science, Bangalore, Karnataka, India. 560012.}
\author{Rajib Biswas}
\email{rajib@iittp.ac.in}
\affiliation[IITT]
{Department of Chemistry, Indian Institute of Technology Tirupati, Yerpedu, Tirupati, Andhra Pradesh, India. 517619.}

\title{Azurin-Based Peptide p28 Arrests the p53-HDM2 Interactions: A Novel Anti-Cancer Pathway}


\begin{document}

\begin{abstract}
\onehalfspacing
Azurin and its derived peptides, particularly p28, have shown strong anticancer effects, primarily due to their capacity to stabilize the p53 tumor suppressor protein and prevent its degradation. Previous studies have demonstrated that p28 interacts with the DNA-binding domain of p53, protecting it from degradation mechanisms. Building on these insights, our research explored whether the p28 peptide influences other cancer-related pathways beyond just p53 stabilization. Specifically, we focused on the interactions between the p28 peptide and Human Double Minute 2 (HDM2), a protein known to negatively regulate p53’s tumor-suppressive activity.

HDM2 plays a critical role in downregulating p53 by binding to its transactivation domain (TAD), leading to a reduction in p53’s ability to perform its tumor-suppressive functions. Our study aimed to determine if p28 disrupts the interaction between HDM2 and p53, and if so, to elucidate the molecular mechanisms behind this effect. Using HADDOCK docking combined with molecular dynamics simulations, we identified three stable conformations of the HDM2-p28 complex. These conformations effectively occupy HDM2's hydrophobic pocket, allowing for sustained inter-chain interactions and exhibiting favorable binding energies. Through further analysis, we pinpointed the essential residues involved in these interactions and calculated their interaction energies using the Molecular Mechanics Poisson-Boltzmann Surface Area (MMPBSA) method.

Our findings reveal that by blocking HDM2’s binding sites, the p28 peptide can help maintain p53’s transcriptional activity, thereby enhancing its tumor-suppressive roles, including the induction of apoptosis and the arrest of the cell cycle in cancer cells. This study significantly advances our understanding of azurin-derived peptides' anticancer mechanisms, demonstrating the therapeutic potential of p28 as a promising peptide-based agent for cancer treatment. Furthermore, these findings may open doors to designing additional peptide-based therapies targeting HDM2 and other pathways critical for cancer progression, offering a new avenue in the development of effective anticancer therapeutics.
\end{abstract}

% \twocolumn
% \setlength{\textwidth}{7in}
% \setlength{\columnsep}{40pt}

% \justify
\section{Introduction}
\onehalfspacing
Human double minute 2 (HDM2), the human homologue of the originally identified mouse double minute 2 (MDM2), is a 491-amino acid phosphoprotein that plays a critical role in regulating the p53 tumor suppressor protein.\cite{11MDM2-1987} The p53 tumor suppressor is a pivotal regulator of cellular stress responses, playing a crucial role in enforcing cell growth arrest, inducing senescence, and initiating apoptosis in response to various forms of cellular stress.\cite{p531997,p53Wu1997} Its ability to suppress the propagation of cells harboring damaged or potentially mutagenic DNA is vital for maintaining genomic integrity. The p53 protein mediates these protective effects primarily through a transcription-dependent mechanism.\cite{12mdm21992,p53trans2002} A central feature of p53 biology is its tight regulation at the protein level, as cellular p53 concentrations are a key determinant of its function. In unstressed cells, p53 is maintained at low levels due to its rapid degradation, with a half-life of 5 to 30 minutes, primarily controlled by the ubiquitin ligase MDM2.\cite{p531997,p53Wu1997} Upon exposure to stress, p53 levels are swiftly elevated, allowing it to exert its tumor suppressor functions effectively.

\begin{figure*}[htb!] \center
\includegraphics[width=0.7\textwidth]{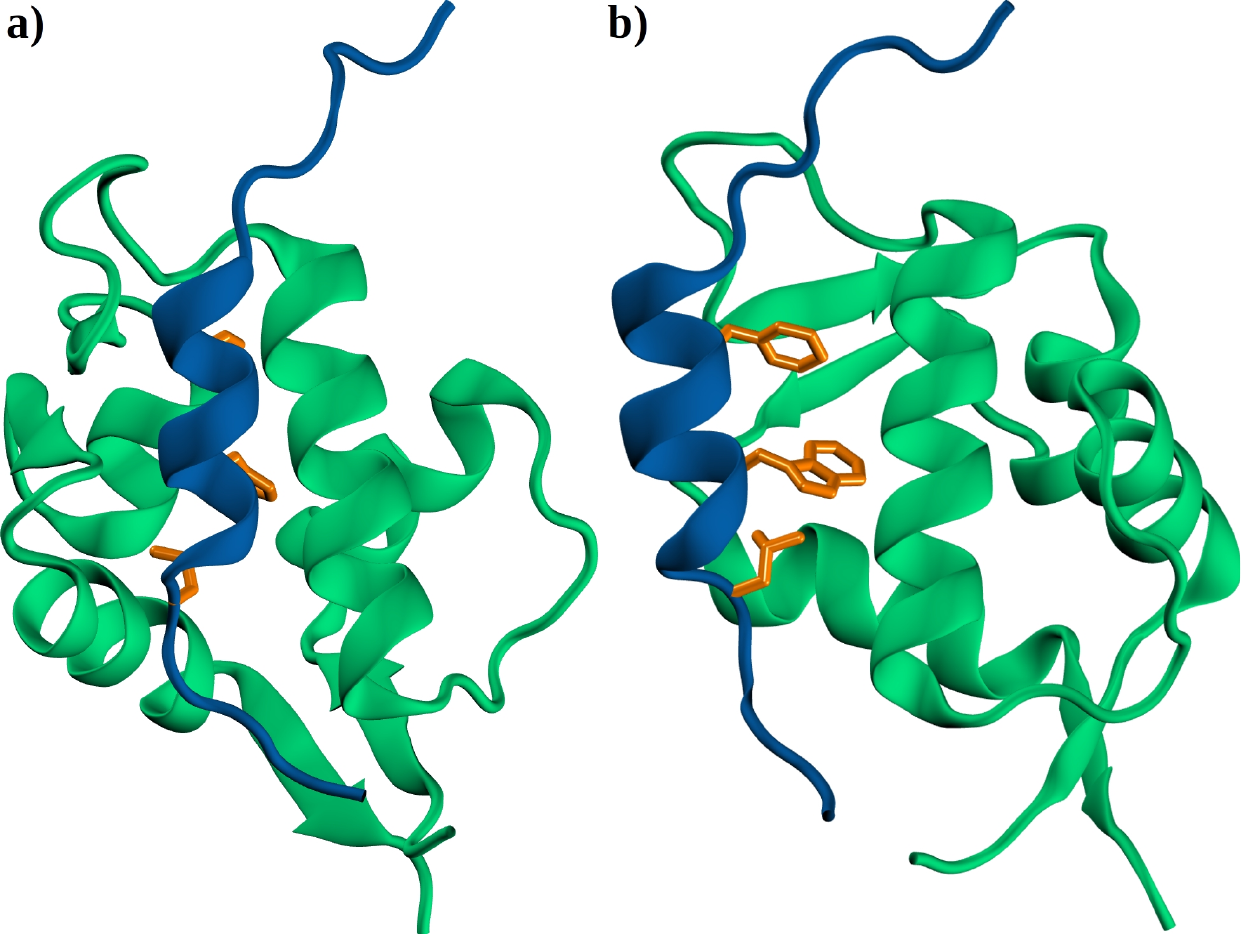}
\caption{\label{fig6.1} (a) Cartoon representation of the native HDM2-p53 interaction sites and the hydrophobic cleft. p53 (residues 10-30) is colored blue and HDM2 (residues 25-111) in green. (b) Residues PHE19, TRP23, and LEU26 of p53, represented in stick model and colored orange, binding to the hydrophobic pocket of HDM2.}
\end{figure*}

Structural studies of the p53-MDM2 complex reveal that MDM2’s amino-terminal domain binds to p53’s transactivation domain (TAD) within a hydrophobic pocket.\cite{Kussie1996} This interaction involves residues 25–109 of MDM2 forming the HDM2 NTD and a 15-amino acid amphipathic peptide in p53, with key binding sites located at residues 18–26 of p53.\cite{Chen-1993,Bottger1996a,BOTTGER1997} Residues PHE19, TRP23, and LEU26 of p53, which bind to the hydrophobic pocket of HDM2, play crucial role for this protein-protein interaction. These interactions are essential for the regulation of p53's tumor suppressor activity, as HDM2 binding leads to its functional inhibition.\cite{Chen-1993,Oliner1993} The interaction is characterized by a precise key-lock fit, with p53’s amphipathic $\alpha$-helix fitting into a hydrophobic cleft in MDM2. THR18 in p53 stabilizes this $\alpha$-helix, while MDM2’s binding cleft, formed by residues 54-74 and 94-104, features a groove lined with hydrophobic and aromatic residues. Conversely, mutations at GLY58, GLU68, VAL75, or CYS77 in MDM2 can disrupt its binding to p53.\cite{Freedman1997} Figure \ref{fig6.1} illustrates the native HDM2-p53 interaction sites, highlighting the hydrophobic cleft and the crucial binding residues of p53. The binding affinity of p53-MDM2 complex ranges from 60 to 700 nM, depending on the number of residues of the p53 protein involved.\cite{LAI2000,SAKAGUCHI2000,SCHON2002} Such binding inhibits p53's ability to activate transcription by physically blocking its interaction with the transcriptional machinery.

Phosphorylation also plays a key role in regulating this interaction. While phosphorylation of SER15 and SER20 does not affect p53-MDM2 binding, phosphorylation of THR18 weakens the interaction by 10-fold, indicating that THR18 phosphorylation is critical for disrupting p53-MDM2 binding.\cite{SAKAGUCHI2000,SCHON2002} DNA damage-induced disruption of this complex \textit{in vivo} specifically requires THR18 phosphorylation, though the stabilization of p53 after ionizing radiation involves a phosphorylation cascade beginning with SER15, which facilitates subsequent phosphorylation of THR18. Nuclear magnetic resonance studies reveal that p53 binding induces significant conformational changes in MDM2 that extend beyond the binding cleft. Overall, conformation and hydrophobicity are key determinants of the p53-MDM2 interaction.

As an E3 ubiquitin ligase, MDM2 is primarily responsible for tagging p53 for degradation, thus controlling its cellular levels.\cite{Haupt1997,Kubbutat1997} MDM2 features a self- and p53-specific E3 ubiquitin ligase activity located within its evolutionarily conserved COOH-terminal RING finger domain, which is essential for its function. This domain is responsible for transferring ubiquitin to lysine residues in the COOH terminus of p53, marking it for degradation. MDM2 and p53 regulate each other through an autoregulatory feedback loop which is essential for cellular homeostasis and effective response to various cellular stress.\cite{Chen-1993,mdm2-1993,Picksley-1993} p53 induces the expression of MDM2, which in turn inhibits p53 by promoting its degradation and nuclear export. This feedback mechanism ensures that p53 activity is tightly controlled. Also, various DNA-damaging agents and oncogenes can activate p53 by inducing phosphorylation of both p53 and MDM2, leading to disruption of their interaction and stabilization of p53, thereby allowing it to perform its tumor-suppressive functions.

Due to the well-characterized structural and biological nature of the hydrophobic interaction between p53 and MDM2, designing small lipophilic molecules that disrupt or inhibit this interaction has become a promising therapeutic approach. Notably, the MDM2 protein presents structurally well-defined binding sites, whereas p53 undergoes dynamic conformational changes upon phosphorylation. This suggests that inhibitors should more closely mimic p53 rather than MDM2. Additionally, the p53 interface consists of a single short contiguous sequence of amino acids, making it a suitable target for small-molecule inhibitors. The interaction between p53 and MDM2 involves only three hydrogen bonds, with the most critical bond being formed by TRP23 on p53. The relatively small contact surface between these two proteins further supports the idea that small inhibitory peptides or molecules could effectively disrupt their interaction.

Previous studies has suggested that azurin and its derived peptides possess significant anticancer properties, demonstrating the ability to inhibit the growth and proliferation of various cancer cell types.\cite{p53Valentina2007,cancerp28,Huang2020} Some works have shown that the p28 peptide (a small peptide derived from azurin having residues LEU50-ASP77 of azurin) can stabilize the p53 tumour suppressor protein, preventing its degradation and thereby arresting cancer cell growth.\cite{Taylor2009,Yamada2013} By binding to the DNA-binding domain (DBD) of p53, p28 effectively shields it from cellular mechanisms that lead to its degradation.\cite{Bizzarri2011,SIGNORELLI2017,BIZZARRI2019,Hu2023} Another study has shown that the binding of azurin to p53 significantly decreases the association rate constant and binding affinity between MDM2 and p53, without blocking MDM2's access to p53's binding pocket.\cite{Fabio2011} Building on these findings, our study aims to investigate the molecular interactions between the azurin-based peptide p28 and the cancer-related protein HDM2. Specifically, we seek to uncover the mechanisms by which azurin exerts its anticancer effects, particularly through disrupting the HDM2-p53 interaction, a critical pathway in tumor suppression. By focusing on the binding affinity and conformational dynamics of these complexes, we provide deeper insights into the therapeutic potential of p28 as a promising peptide-based anticancer agent, contributing to the development of novel treatments targeting HDM2-associated cancers.

\section{Methods}
As mentioned earlier, HDM2 acts as a downregulator of p53 by binding to its transactivation domain (TAD) which diminishes its tumor-suppressive function.  In this study, we investigate the interaction between the p28 peptide and the N-terminal domain (NTD) of HDM2, the region responsible for binding to the TAD of p53, which leads to p53 inhibition. Our approach includes protein docking studies, molecular dynamics simulations, and energy calculations to evaluate the binding affinity and stability of the HDM2-p28 complex.

\subsection{Docking of HDM2 and p28 Peptide}
We used the N-terminal domain (NTD) of HDM2 with a resolution of 1.8 Å obtained from the RCSB Protein Data Bank (PDB ID: 6Q96) and extracted the p28 peptide fragment from the azurin structure (PDB ID: 4AZU).\cite{6q96,azu4} Both the HDM2 NTD and the p28 peptide were equilibrated at 300 K and 1 bar pressure for 100 ns within the NPT ensemble. The final equilibrated structures of HDM2 NTD and p28 peptide was then used for docking studies. Docking was performed using the HADDOCK (High Ambiguity Driven protein-protein DOCKing) server, which generated the top 10 most reliable clusters based on energy scores.\cite{hdock2,haddock2024}

The HADDOCK methodology integrates experimental data with computational docking, allowing for the refinement of biomolecular complexes through a multi-stage approach. In this study, the docking process followed a systematic approach to capture the interaction between the p28 peptide and the N-terminal domain (NTD) of HDM2. A key feature of HADDOCK is the use of Ambiguous Interaction Restraints (AIRs), which convert experimental data, such as NMR or mutagenesis results, into distance restraints. These restraints are incorporated into the energy function, guiding the docking calculations toward more accurate models.

The first step in the HADDOCK docking process involves randomization of orientations and rigid-body minimization. the interacting partners, such as the p28 peptide and HDM2 NTD, are treated as rigid bodies, meaning that all geometrical parameters (bond lengths, bond angles, and dihedral angles) remain fixed. The molecules are separated in space and randomly rotated about their centers of mass to explore possible configurations. Next, a rigid-body energy minimization step is performed, allowing the partners to rotate and translate in order to optimize their interaction, thereby identifying low-energy conformations for subsequent refinement. The second stage of HADDOCK introduces flexibility to the interacting partners through a molecular dynamics-based refinement process, enhancing interface packing. Initially, bond lengths and angles remain fixed, while only the molecular orientations are optimized. Flexibility is then applied to the interface, defined by intermolecular contacts within a 5 Å distance, allowing for a more accurate representation of the interactions at the binding site. Next, the side chains of the residues in this region are allowed to relax in a second refinement step. Finally, both backbone and side-chain flexibility are introduced, enabling more accurate conformational adjustments and a better representation of the interaction at the binding site.

In the final phase of the docking, the complex is immersed in a solvent shell to improve interaction energetics. The models undergo a brief molecular dynamics simulation at 300K, with position restraints on the non-interface heavy atoms. These restraints are gradually relaxed, allowing the side chains to adjust and optimize the interaction further.

HADDOCK employs a scoring system based on a combination of factors including van der Waals forces, electrostatics, desolvation energies, and empirical restraints derived from experimental data or predictions. In our analysis, the server generated a set of docking models, grouped into clusters based on similarity. The top 10 clusters generated are ranked according to their HADDOCK and Z-scores, reflecting the reliability and quality of each predicted binding mode.

\subsection{MD Simulations of Docked Structures}
Each of these structures was placed in a simulation box at a distance of 1.2 nm from the box walls and solvated with water. Na$^+$ and Cl$^-$ ions were added to neutralize the system and simulate physiological salt concentrations. We employed the CHARMM36m force field parameters for the protein chains and the TIP3P model for water molecules. Molecular dynamics simulations were carried out using the GROMACS (GRoningen MAchine for Chemical Simulations) software suite, a robust and highly efficient tool that is extensively utilized for advanced molecular dynamics simulations.\cite{abraham2015}

Energy minimization was performed using the steepest descent method to resolve steric clashes, achieving convergence once the forces reached below 1000 kJ/mol/nm. The energy-minimized structure was then equilibrated at 300 K in the NVT ensemble for 1 ns, followed by pressure equilibration at 1 bar in the NPT ensemble for 1 ns. The equilibrated system was then subjected to a production run in the NVT ensemble for 200 ns, with data collected every picosecond. Temperature and pressure were maintained at 300 K and 1 bar, respectively, using the Nosé-Hoover thermostat\cite{nose,hoover} with a relaxation time of 0.5 ps and the Parrinello-Rahman barostat\cite{Parrinello1981, Nose1983} with a relaxation time of 1.0 ps. All van der Waals interactions were truncated at a 1.2 nm cut-off, with the neighbor list updated every ten steps, and electrostatic interactions were calculated using the PME (Particle-Mesh Ewald) method\cite{pme} with the same cut-off. The LINCS (LINear Constraint Solver) algorithm\cite{lincs} was used to constrain the bond lengths of hydrogen atoms. The trajectory was analyzed to evaluate the sustained interactions between HDM2 and p28. We found 3 structures that adapt different binding poses and in which the p28 peptide effectively binds and masks the binding pocket on HDM2 protein.

\subsection{Subsequent Docking Studies with p53 TAD}
The final structures of the above 3 clusters were subjected to a second docking study involving p53 TAD (residues 10-30) obtained from the human p53 structure in the UniProt database (ID: P04637). This docking was performed to study the binding of p53 TAD to the HDM2-p28 complex. The top clusters from these docking studies were analyzed to understand how the presence of the p28 peptide influences the binding modes of HDM2 with p53 TAD.

\section{Results and Discussion}
\subsection{HADDOCK Scores of p28 Docking to HDM2}
\begin{figure*}[htb!] \center
\includegraphics[width=1.0\textwidth]{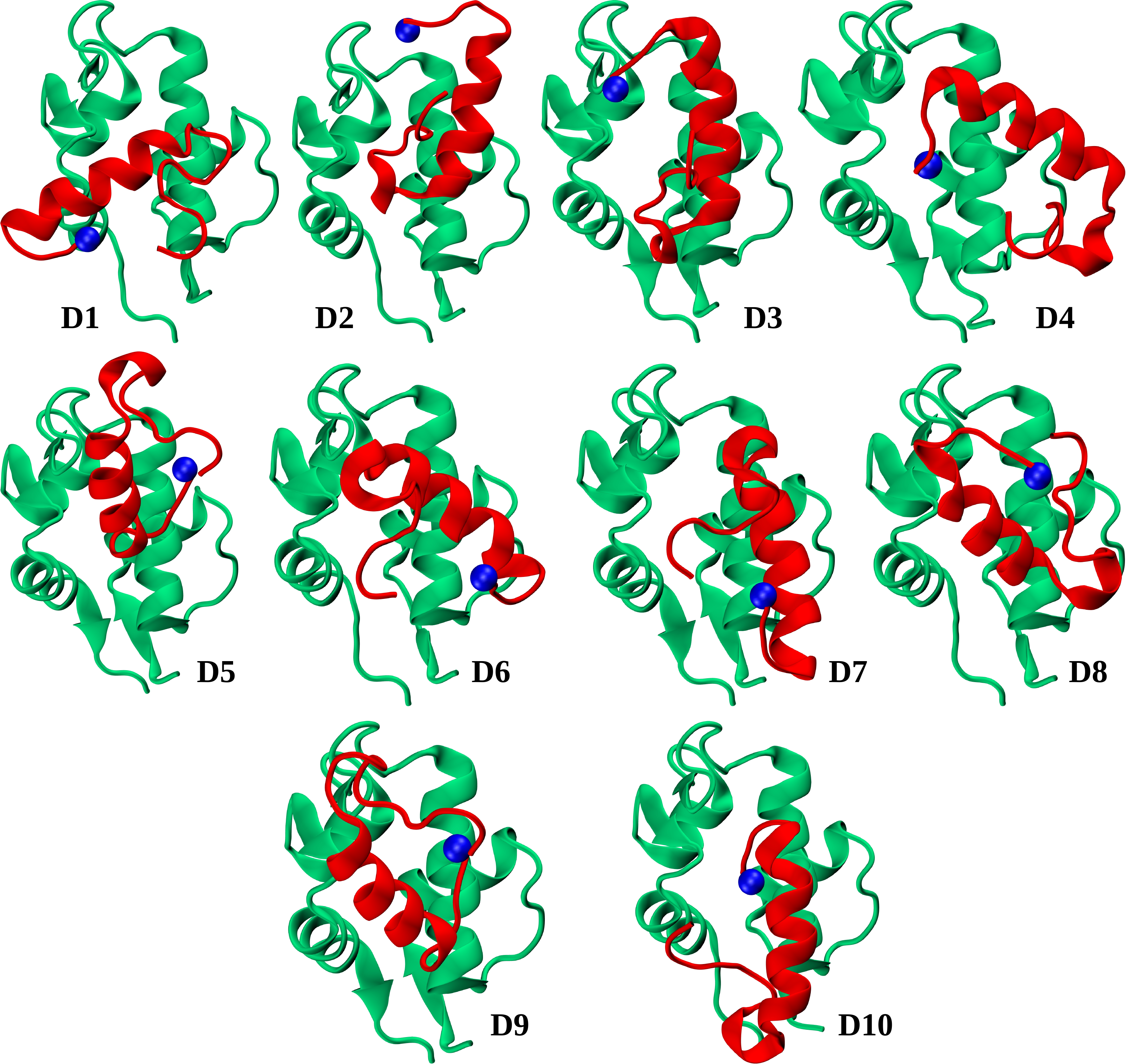}
\caption{\label{fig6.0} Cartoon representation of the docked complexes D1 to D10 obtained from the HADDOCK server showing different binding poses of the p28 peptide with the HDM2 NTD. The HDM2 protein is colored in purple and yellow and the p28 peptide is shown in red. The N-terminal nitrogen atom of the p28 peptide is depicted as a blue sphere.}
\end{figure*}

The docking of the p28 peptide with the N-terminal domain (NTD) of HDM2 yielded a total of 10 distinct cluster structures, highlighting various potential binding modes. In the rest of this article, we refer to these structures as D1, D2, \ldots, D10. The best models from these top 10 HADDOCK clusters are presented in Figure \ref{fig6.0}. Interestingly, in many of these docked complexes, the p28 peptide was found to bind near the region of the hydrophobic pocket of HDM2 NTD. This hydrophobic pocket of HDM2 is known for its interaction with the transactivation domain (TAD) of the p53 tumor suppressor protein, suggesting that p28 may potentially interfere with or mimic this critical binding interface. (Native HDM2-p53 interaction sites and the hydrophobic cleft are depicted in Figure \ref{fig6.1}.) These structures are ranked by the HADDOCK server based on the HADDOCK scores, which quantify the overall docking energy as a weighted sum of various interaction energy terms, and are presented in Table \ref{table6.1}. The HADDOCK score is computed using the following weighted energy equation.
\begin{equation}
\begin{split}
\text{HADDOCK Score} = 1E_{\text{vdw}} + 0.2E_{\text{elec}} + 1E_{\text{desolv}} + 0.1E_{\text{AIR}}\label{eq6.1}
\end{split}
\end{equation}

\begin{table}[ht!]
\centering
  \caption{Details of the clusters generated from the HADDOCK server for HDM2-p28 docking.}
  \label{table6.1}
  \begin{tabular}{ll}
    \hline
    Complex  & HADDOCK Score  \\
    \hline
    D1  & -77.6 +/- 11.0 \\
    D2  & -71.5 +/- 7.5  \\
    D3  & -68.7 +/- 2.0  \\
    D4  & -64.9 +/- 5.5  \\
    D5  & -64.0 +/- 9.2  \\
    D6  & -61.9 +/- 6.2  \\
    D7  & -58.3 +/- 8.0  \\
    D8  & -58.0 +/- 5.6  \\
    D9  & -57.8 +/- 8.6  \\
    D10 & -55.3 +/- 4.6  \\
    \hline
  \end{tabular}
\end{table}

\subsection{Structural Stability of HDM2-p28 Complexes}
These structures were subjected to MD simulations at 300 K for 200 ns to assess the stability of the interactions and to accommodate any conformational changes leading to the most favorable low-energy binding interactions. These MD trajectories of the different equilibrated structures were analyzed to evaluate the stability of various binding poses, sustained interactions and structural behavior of the protein chains. First, the distance between the center of mass (COM) of the protein chains of HDM2 and p28 was calculated to determine whether the p28 peptide continuously interacts with HDM2 throughout the simulations. We found that in all 10 structures, p28 consistently interacted with HDM2, maintaining proximity throughout the simulation. Next, we assessed the temporal evolution of the backbone root mean square deviation (RMSD) and the radius of gyration (Rg) for both protein chains. HDM2 NTD exhibited minimal deviations compared to the initial structure, indicating overall structural stability. The p28 peptide, on the other hand, predominantly retained its folded $\alpha$-helical structure, although slight structural melting was observed in a few cases. We further calculated the solvent accessible surface area (SASA) of the residues forming the hydrophobic pocket of HDM2 to identify which structures had the HDM2 binding site masked by p28. Among all trajectories, the D1 trajectory showed slightly lower SASA values, indicating more extensive masking of the hydrophobic pocket by p28 in this model.
\begin{figure*}[htb!] \center
\includegraphics[width=1.0\textwidth]{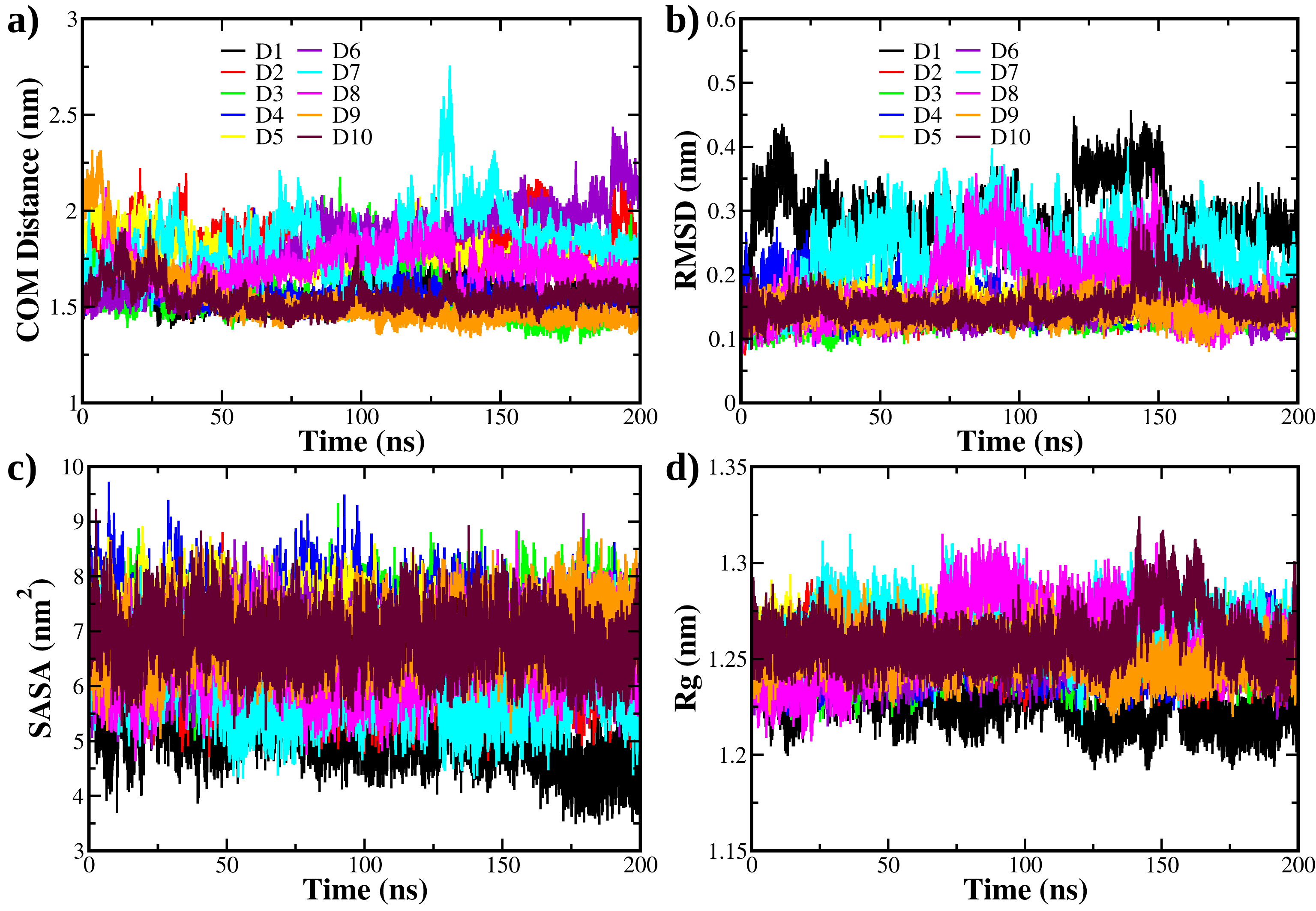}
\caption{\label{fig6.2} (a) Temporal evolution of the distance between the center of mass of HDM2 and p28 protein chains for the best models from the top 10 clusters generated by the HADDOCK server. (b) Changes in the RMSD of the backbone atoms of the HDM2 NTD for these structures in the 200 ns trajectory. (c) Variations in the solvent accessible surface area of residues forming the hydrophobic pocket of HDM2 for the top 10 structures. (d) Temporal evolution of the radius of gyration of the backbone atoms of HDM2 NTD during the simulation.}
\end{figure*}

We further analyzed the root mean square fluctuations of C$\alpha$ carbon atoms (C$\alpha$-RMSF) of individual residues in both protein chains for the last 50 ns of the MD trajectories. The analysis showed that most of the residues of HDM2 remained structurally rigid, closely resembling their native states under the simulated conditions. The p28 peptide exhibits slightly greater fluctuations, as expected, reflecting its considerable structural flexibility. This flexibility may contribute to its ability to adapt to various conformational states, allowing for dynamic interactions with HDM2. In general, residues 5-25 of p28 remain significantly stable in their native $\alpha$-helical structure, probably with their interactions with HDM2. Only the N- and C-terminal residues exhibited higher fluctuations. The p28 peptide in the D7 trajectory showed the highest fluctuations, where the peptide was positioned away from HDM2 and partially unfolded. Similarly, in the D3 trajectory, the p28 peptide exhibited significant structural melting.
\begin{figure*}[htb!] \center
\includegraphics[width=1.0\textwidth]{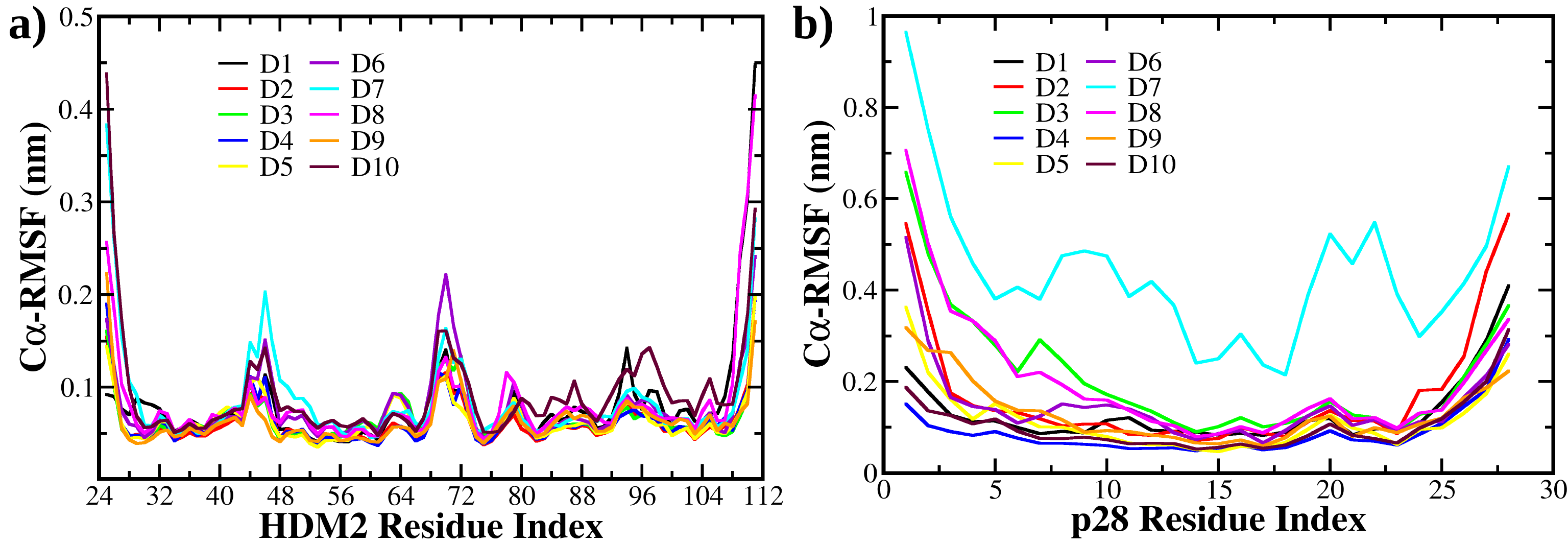}
\caption{\label{fig6.3} Root mean square fluctuations (RMSF) of C$\alpha$ carbon atoms of (a) HDM2 protein and (b) p28 peptide obtained from the last 50 ns of the MD trajectories of different docked structures. The p28 peptide shows higher fluctuations, indicating its large structural flexibility.}
\end{figure*}

\subsection{Conformational Diversity of HDM2-p28 Complexes}
\begin{figure*}[htb!] \center
\includegraphics[width=1.0\textwidth]{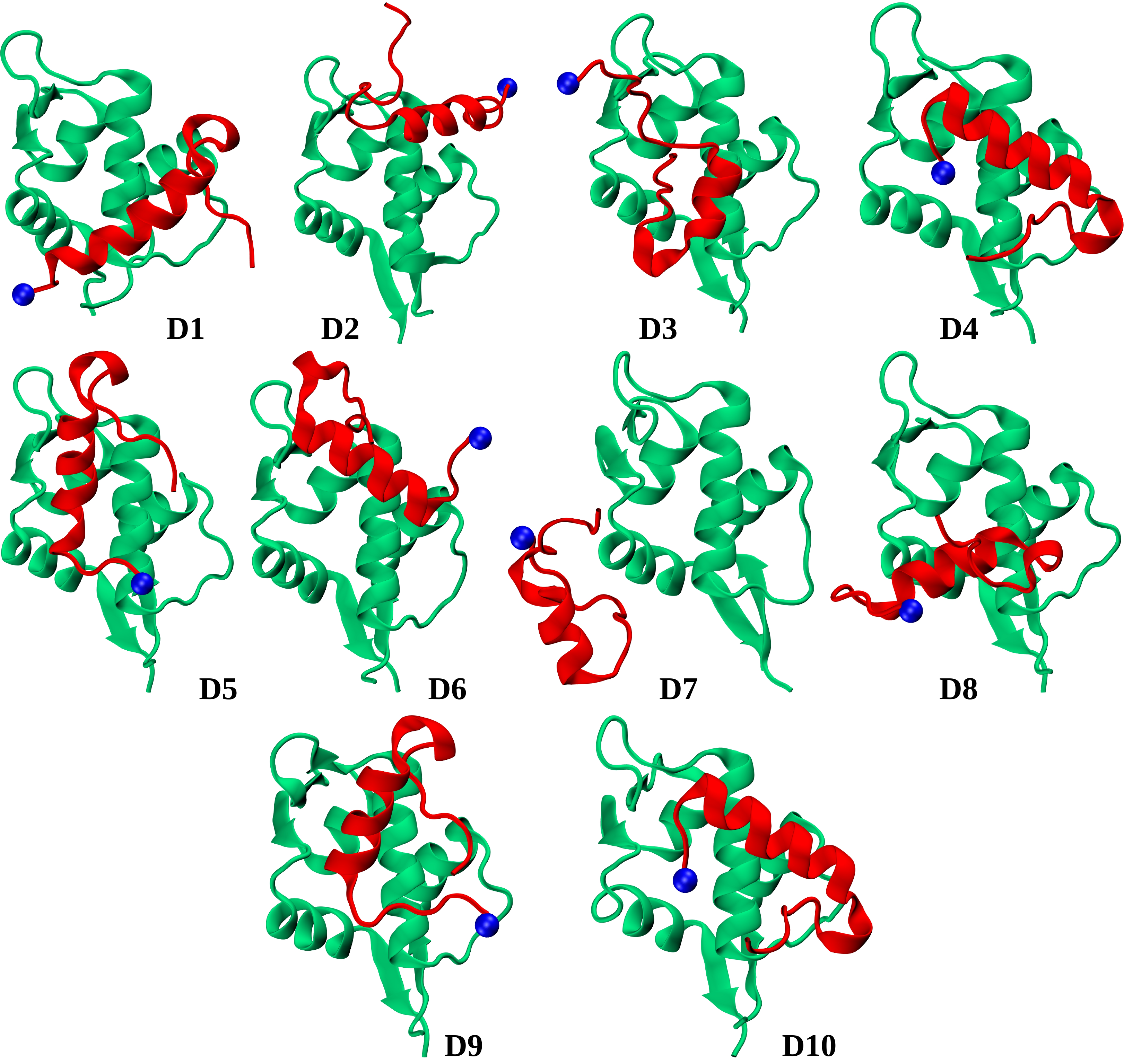}
\caption{\label{fig6.4} Cartoon representation of the final binding poses of the p28 peptide with HDM2 after 200 ns of simulation for docked complexes D1 to D10. The HDM2 protein is colored in purple and yellow, while the p28 peptide is shown in red. The N-terminal nitrogen atom of the p28 peptide is depicted as a blue sphere.}
\end{figure*}

After the 200 ns simulation, several docked complexes converged to similar binding modes. Three pairs of such similar complexes were identified: D2-D6, D4-D10, and D5-D9. Upon analyzing the binding regions in the final structures, we observed that in D1 and D2, the p28 peptide binds near the active region of HDM2. However, in D2, the active site remains accessible. A similar p28 conformation is observed in D6, where the active site is still open. In D3, the p28 peptide is partially unfolded, failing to block the active region effectively. In D4, the N-terminal end of p28, with its hydrophobic leucine residue, interacts with the hydrophobic pocket of HDM2. This binding pattern is also seen in D10. In D5 and D9, p28 aligns perfectly within the hydrophobic groove. In D7, the p28 peptide moved away from the active site during the simulation, while in D8, it effectively blocked the active site. Now, from these 10 complexes, we narrowed the selection to three conformationally relevant structures that effectively block the hydrophobic pocket, display structurally unique binding patterns, maintain sustained inter-chain interactions, and exhibit favorable binding energies. These are complexes D5, D8, and D4, which are further studied in-depth for their various aspects of binding characteristics. The final binding poses of the p28 peptide with HDM2 are depicted in Figure \ref{fig6.4}.

\subsection{Contact Frequency Maps}
\begin{figure}[htb!]\center
\includegraphics[width=1.0\linewidth]{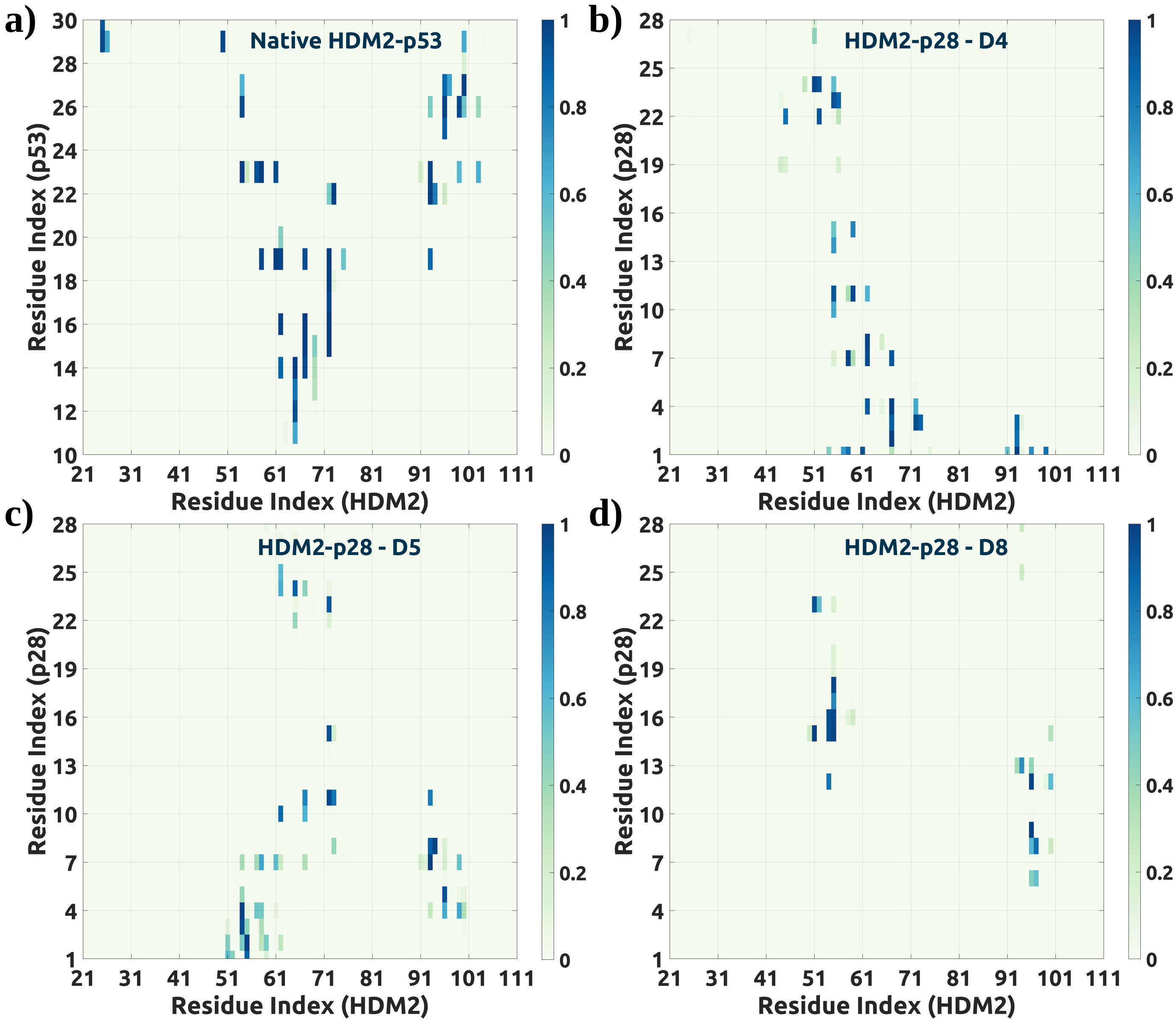}
\caption{\label{fig6.5}(a) Contact frequency map of the native interaction between HDM2 and p53 residues. Contact frequency map of the HDM2-p28 interactions for (b) complex D4, (c) complex D5 and (d) complex D8. The map highlights prominent and sustained interactions between HDM2 and p53 residues, revealing key interaction points that inhibit the functional activity of p53.}
\end{figure}

In these three complexes, we calculated the contact frequency between interacting residues, which is depicted as a contact frequency map in Figure \ref{fig6.5}. The contact frequency map represents the probability of contacts between residues of HDM2 and p28, using data from the last 50 ns of the MD trajectories. A contact between two residues is defined if the distance between any of their heavy atoms is less than 4.5 Å. These contact maps provide valuable insights into the residues that engage in favorable and sustained interactions between HDM2 and p28. Contact frequency maps are helpful for understanding protein-protein interactions, as they highlight key residues involved in stable binding interfaces. By analyzing these maps, we identified crucial interaction hotspots between HDM2 and p28 and assessed how p28 modulates the binding region of HDM2, potentially impacting its interaction with other cell growth regulatory proteins such as p53. The high contact frequency regions suggest critical sites for the stabilization of the HDM2-p28 complex.

To compare the contacts between the p28 peptide and HDM2 with the native p53-HDM2 interactions, we also mapped the contact frequency of the p53-HDM2 binding interface using the last 10 ns trajectory of a 100 ns simulation of this complex at 300 K. The key residues that form the native HDM2-p53 interactions are predominantly hydrophobic, along with some other residues involved in hydrogen bonding interactions. The region of the HDM2 N-terminal domain (NTD) involved in these native interactions includes key residues within the range of 54-67, as well as residues 50, 72, 93, 96, and 99. Important hydrophobic residue pairs include LEU54-TRP23, ILE61-PHE19, VAL93-LEU22, VAL93-TRP23, MET62-PHE19, ILE99-LEU26, LEU54-LEU26, ILE61-TRP23, and LEU57-TRP23, where the first residue is from HDM2 and the second from p53. In addition, residue pairs such as TYR67-SER15, GLN72-THR18, GLN72-GLN16, GLN72-SER15, MET62-GLN16, and GLU25-ASN29 are involved in hydrogen bonding and polar interactions.

\begin{figure}[htb!]\center
\includegraphics[width=1.0\linewidth]{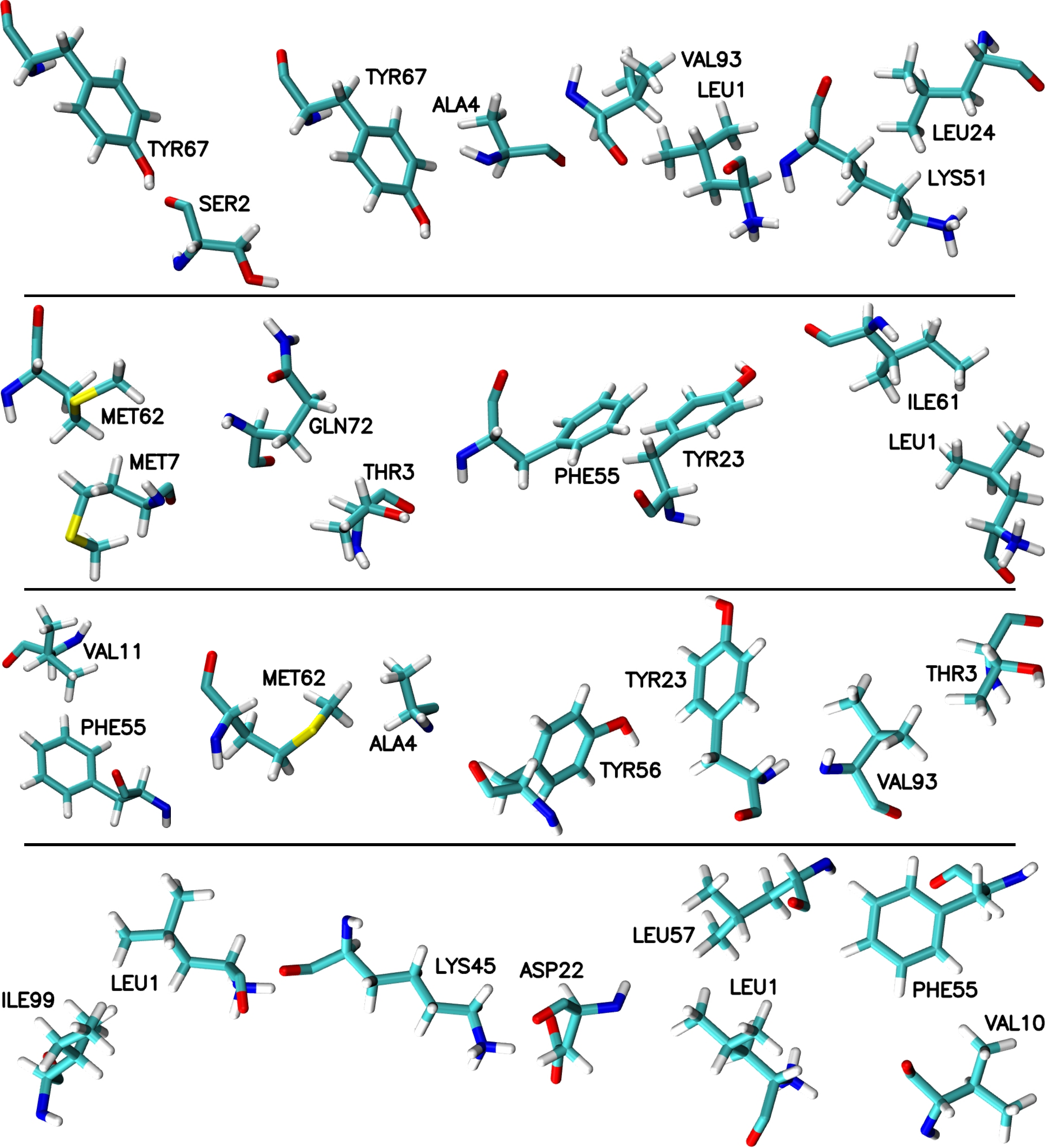}
\caption{\label{fig6.5a}Stick model representation of amino acid pairs forming prominent interactions between HDM2 and the p28 peptide in complex D4.}
\end{figure}

In the D4 docked complex, HDM2 residues 51, 52, 55-62, 67, 73, 93, and many others formed stable interactions during the 200 ns simulation. The p28 peptide residues in the sequence range of 1-4, 7-11, and 22-23 are involved in masking the hydrophobic pocket of HDM2 by interacting with the above HDM2 residues. Important hydrophobic residue pairs include TYR67-ALA4, VAL93-LEU1, MET62-MET7, PHE55-TYR23, ILE61-LEU1, PHE55-VAL11, TYR67-MET7, and MET62-ALA4. Additionally, other notable interactions were observed between residue pairs TYR67-SER2, MET62-GLN8, GLN59-VAL11, LYS51-LEU24, GLN72-THR3, GLU52-LEU24, GLU52-ASP22, TYR67-THR3, and HIS73-THR3. Figure \ref{fig6.5a} depicts the amino acid pairs involved in stable interactions between HDM2 and the p28 peptide in D4 complex.

\begin{figure}[htb!]\center
\includegraphics[width=1.0\linewidth]{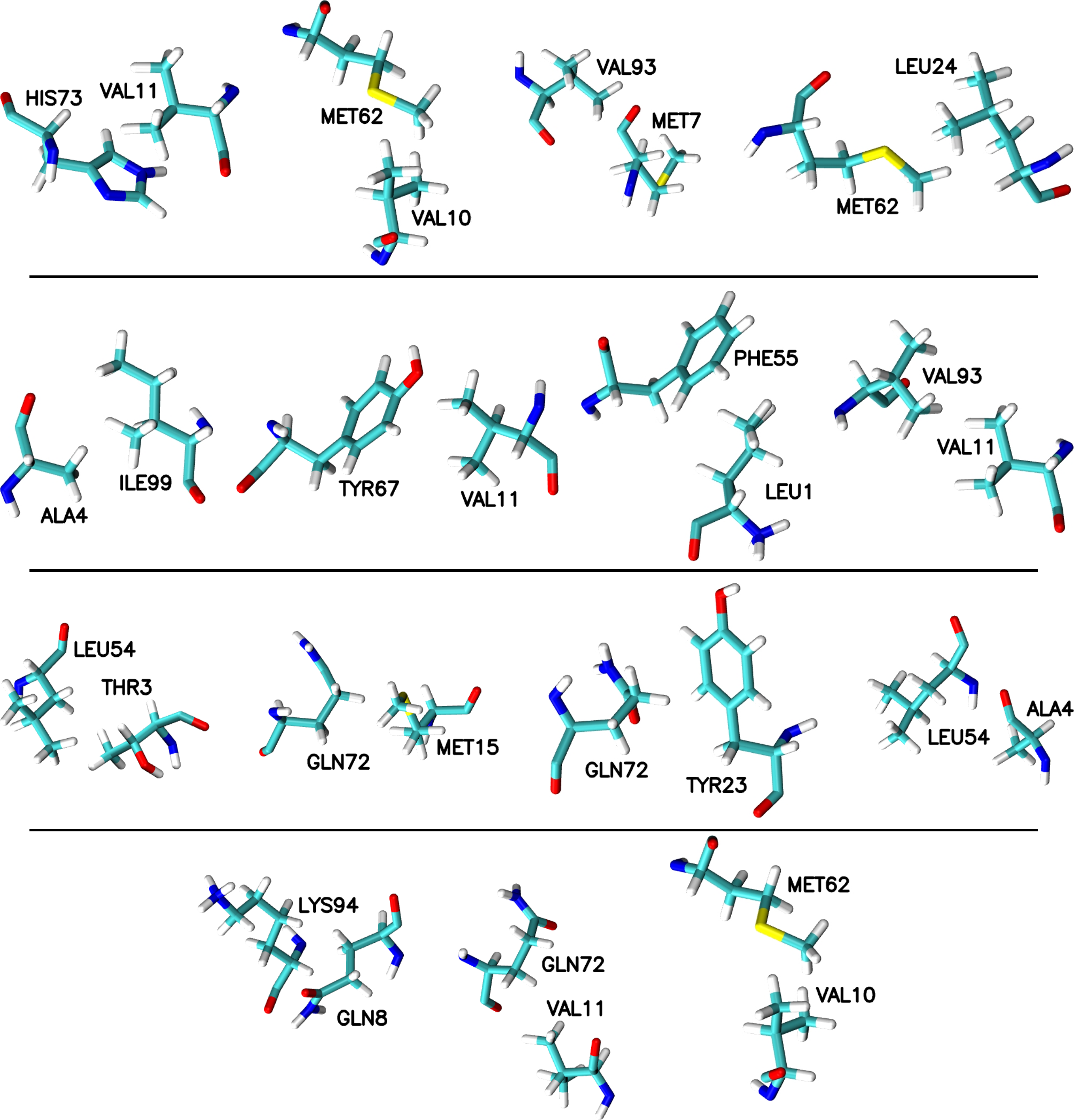}
\caption{\label{fig6.5b}Stick model representation of amino acid pairs forming prominent interactions between HDM2 and the p28 peptide in complex D5.}
\end{figure}

In the D5 complex, the HDM2 residues at the binding interface were similar to those in the D4 complex. In the p28 peptide, residues 1-8, 10, 11, 15, 23, 24, and others were involved in crucial contacts. Important hydrophobic and polar interactions include PHE55-SER2, VAL93-MET7, LEU54-ALA4, LYS94-GLN8, GLN72-MET15, GLN72-TYR23, PHE55-LEU1, MET62-VAL10, VAL93-VAL11, and TYR67-VAL11. Figure \ref{fig6.5b} illustrates the amino acid pairs involved in forming prominent interactions within this complex.

\begin{figure}[htb!]\center
\includegraphics[width=1.0\linewidth]{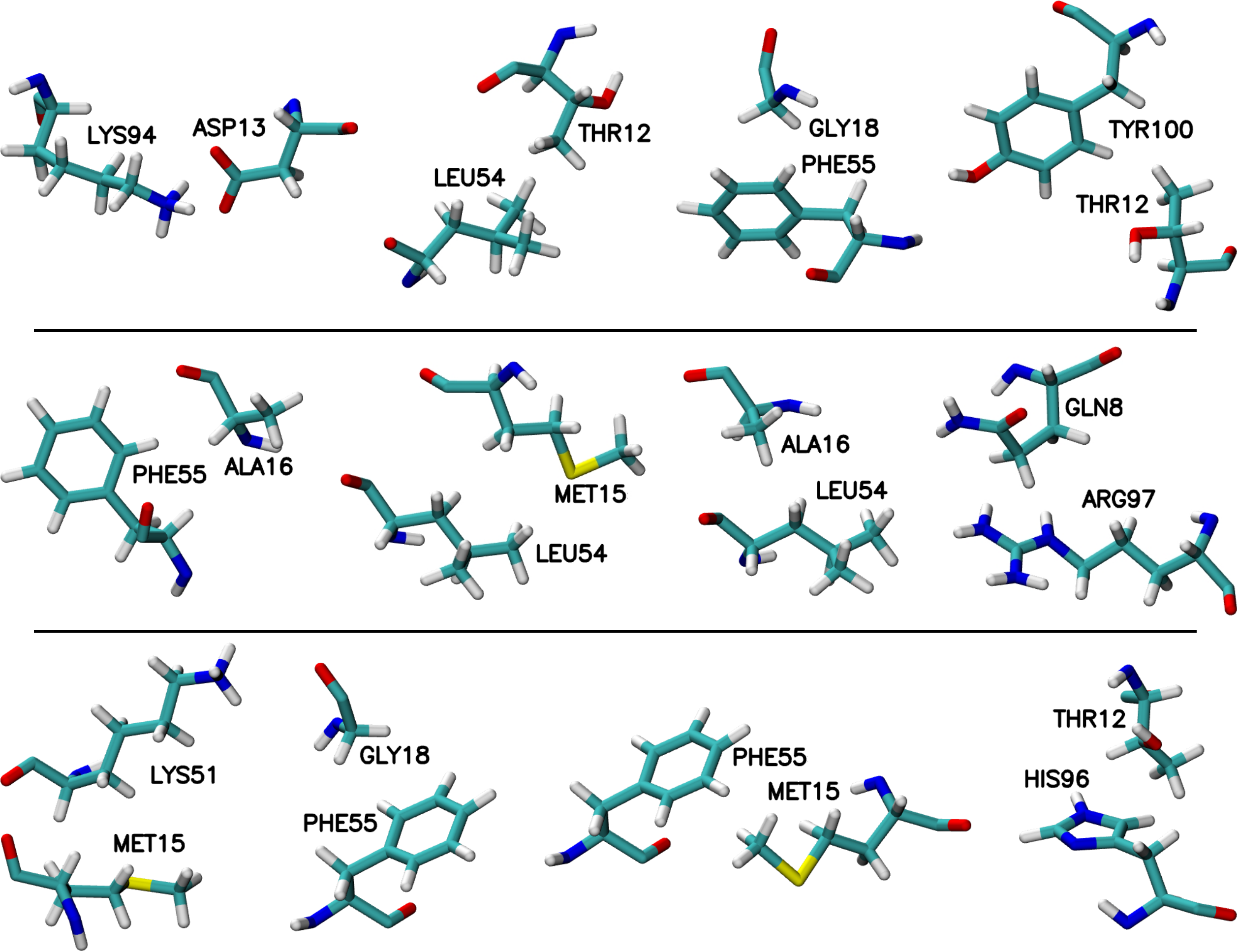}
\caption{\label{fig6.5c}Stick model representation of amino acid pairs forming prominent interactions between HDM2 and the p28 peptide in complex D8.}
\end{figure}

In the D8 complex, during the 200 ns trajectory, HDM2 NTD residues 51, 54, 55, 94, 96, and 97, along with p28 residues 8, 9, 12-18, and 23, contributed to the stability of the complex. Prominent interacting residue pairs included LYS51-MET15, HIS96-GLY9, HIS96-THR12, PHE55-MET15, PHE55-GLY18, PHE55-ALA16, LEU54-MET15, LEU54-ALA16, LYS51-TYR23, ARG97-GLN8, LEU54-THR12, PHE55-SER17, and LYS94-ASP13. In this complex, the number of hydrophobic interactions was comparatively fewer. Figure \ref{fig6.5c} represents the amino acid pairs forming prominent interactions for this complex.

\subsection{Binding Energetics}
Further, the Molecular Mechanics Poisson–Boltzmann Surface Area (MMPBSA) method was employed to evaluate the binding energetics of the HDM2-p28 docked complexes.\cite{mmpbsa2015} This analysis was conducted using the gmxMMPBSA package.\cite{mmpbsa2012,mmpbsa2021} MMPBSA is a widely used approach to estimate binding free energies by combining molecular mechanics (MM) energy terms with solvation energies calculated via the Poisson-Boltzmann (PB) equation and surface area (SA) contributions.

The binding free energy $\Delta G_{\text{binding}}$ was computed as the difference between the free energies of the complex, the receptor (HDM2), and the ligand (p28 peptide) as described in equation \ref{eq6.2}. The total binding energy is derived from the sum of van der Waals, electrostatic, polar solvation, and non-polar solvation components, providing insights into the energetic contributions stabilizing the complexes. The calculations were based on snapshots extracted from the 200 ns molecular dynamics trajectories to capture representative conformations and interactions in each of the docked complexes.
\begin{equation}
\begin{split}
\Delta G_{\text{bind}} = (E_{\text{complex}} - E_{\text{receptor}} - E_{\text{ligand}}) + (\Delta G_{\text{polar}} + \Delta G_{\text{nonpolar}})
\label{eq6.2}
\end{split}
\end{equation}

\begin{figure}[htb!]\center
\includegraphics[width=1.0\linewidth]{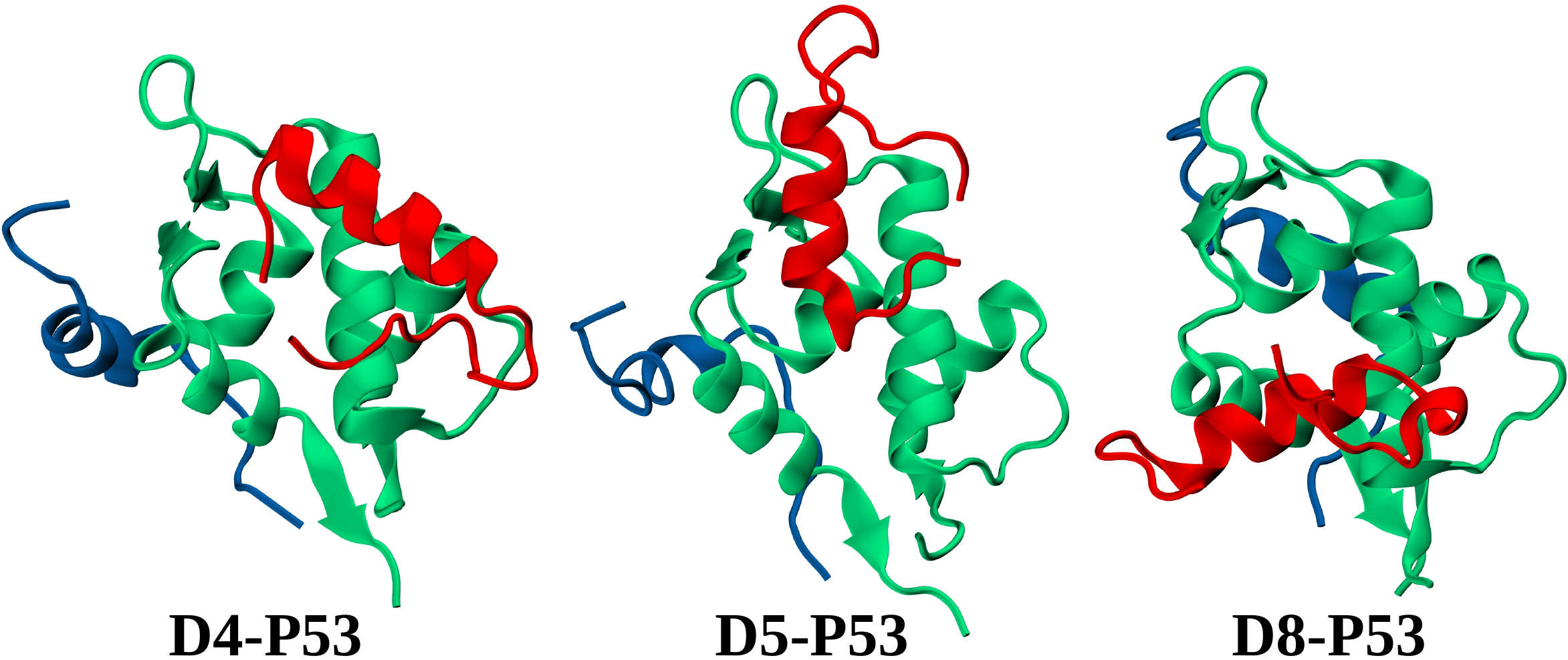}
\caption{\label{fig6.6} Structures of the three relevant complexes from the 200 ns equilibrium MD run docked with p53 TAD, labeled as D4-P53, D5-P53 and D8-P53. These structures indicate that, in the presence of the p28 peptide, HDM2 is unable to interact with the p53 TAD. The HDM2 N-terminal domain (NTD) is colored green, the p28 peptide is shown in red, and the p53 transactivation domain (TAD) is depicted in blue.}
\end{figure}

For the D4 complex, the MMPBSA binding energy was obtained to be $-37.3 \pm 7.87$ kcal/mol, and for the D5 and D8 complexes, it was found to be $-41.92 \pm 4.49$ kcal/mol and $-41.57 \pm 6.91$ kcal/mol, respectively. These highly negative interaction energies indicate a strong affinity of the p28 peptide for the hydrophobic pocket of HDM2 NTD, potentially hindering its association with the p53 protein. The consistency of the binding energies across the complexes suggests that the p28 peptide consistently stabilizes its interaction with HDM2, thereby potentially disrupting the normal binding of HDM2 to its native substrates. This analysis helped to quantify the energetic favorability of the different binding modes observed in the D4, D5, and D8 complexes (Figure \ref{fig6.6}).

\subsection{Docking of HDM2, p28 and p53 TAD Protein Chains}
\begin{figure}[htb!]\center
\includegraphics[width=1.0\linewidth]{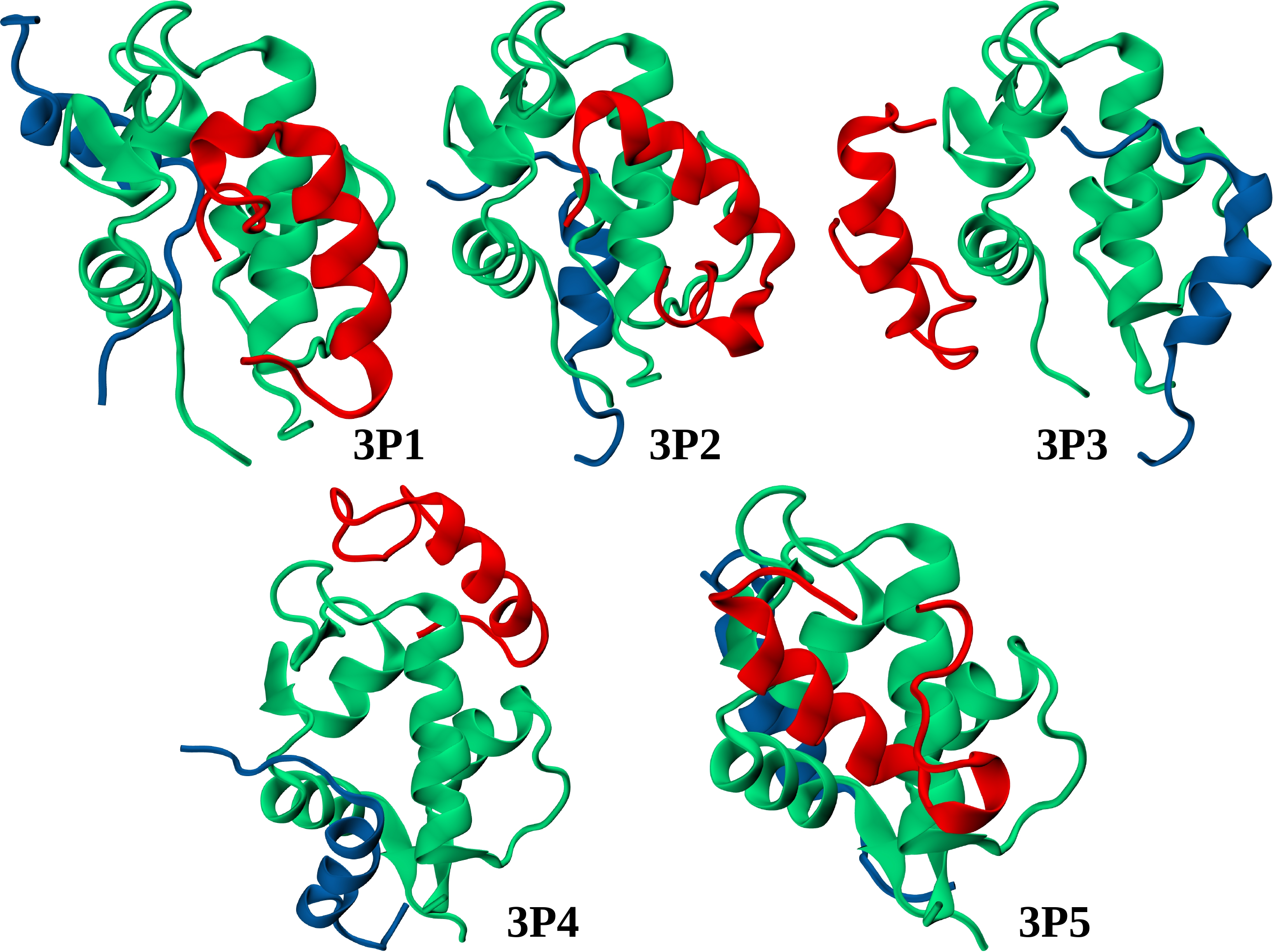}
\caption{\label{fig6.7}First five structures obtained from the docking of the three protein chains - HDM2, p28, and p53 TAD, labeled as 3P1, 3P2, 3P3, 3P4 and 3P5. The structures show that p28 preferentially binds to the HDM2 hydrophobic pocket, preventing HDM2 from interacting with the p53 protein. The HDM2 N-terminal domain (NTD) is colored green, the p28 peptide is shown in red, and the p53 transactivation domain (TAD) is depicted in blue.}
\end{figure}

Further, we explored these different configurations to understand how they interact with p53 TAD by performing a second docking of the HDM2-p28 complex with p53 TAD. From the HADDOCK cluster structures, we observed that due to the presence of the p28 peptide at the HDM2 binding pocket, HDM2 is unable to interact with p53, making only limited contacts between p53 TAD and its own other solvent-exposed regions. We depict snapshots of the best models obtained from this docking, labeled as D4-P53, D5-P53, and D8-P53, in Figure \ref{fig6.6}. This highlights the efficacy of the p28 peptide in preserving the transcriptional activity of the p53 protein. Additionally, we performed another docking of these with all three protein chains simultaneously. The results from the top clusters, scored by HADDOCK, revealed that the p28 peptide consistently binds to the hydrophobic pocket of HDM2, thereby preventing HDM2 from binding to the p53 TAD. The top 5 structures from this docking of the three protein chains, labeled as 3P1, 3P2, 3P3, 3P4 and 3P5 are shown in Figure \ref{fig6.7}.

\section{Conclusions}
This study aimed to understand the mechanisms underlying the suppression of tumor cell growth by the p28 peptide from azurin, focusing on its interactions with the cancer-related protein HDM2. We investigated various possibilities of p28 peptide binding to HDM2 using a combination of docking, molecular dynamics (MD) simulations, and detailed analysis of the resulting docked structures. Our findings reveal that the HDM2 N-terminal domain (NTD) preferentially interacts with the p28 peptide, thereby preventing HDM2 from binding to the p53 protein. This preservation of p53’s transcriptional activity is crucial for regulating cancer cell growth and apoptosis.

From the 200 ns MD simulations of the best docked HDM2-p28 complexes, several structures converged to similar low-energy binding modes. Among these, three complexes, D4, D5, and D8, emerged with distinct binding modes, maintaining sustained inter-chain interactions that effectively block the hydrophobic pocket of HDM2. A detailed analysis of the contact interfaces identified residues 54-72 and 93-99 of HDM2 as critical for its interaction with p53, and the p28 peptide effectively interacts with these residues, potentially hindering HDM2 from binding p53.

We employed the MMPBSA method to evaluate the binding stability of these complexes, which showed favorable binding energies in the range of $-37.0$ to $-42.0$ kcal/mol, indicating strong and stable interactions. Furthermore, a subsequent docking of the final MD structures of the HDM2-p28 complexes with p53 TAD demonstrated that in the presence of p28, HDM2 could not interact with p53. Additional studies by simultaneous docking of HDM2, p28, and p53 TAD confirmed that the p28 peptide preferentially binds to the hydrophobic pocket of HDM2 rather than p53 TAD.

Overall, these studies revealed the molecular basis for the effective anticancer properties of the p28 peptide, highlighting its potential to inhibit HDM2’s interaction with p53 and thereby preserve p53’s tumor-suppressing functions. This insight could be valuable in developing targeted cancer therapies involving peptide-based inhibitors like p28.

\begin{acknowledgement}
RB acknowledges the Indian Institute of Technology Tirupati for its support through the new faculty seed grant (NFSG). AJ and RB acknowledge Indian Institute of Technology Tirupati for the computational facilities. RB also acknowledges SERB-DST for funding through other projects that helped to build part of the essential computational facilities used for the present work.\\
\\
The authors declare no competing interests.
\end{acknowledgement}

\singlespacing
\bibliography{reference}

\end{document}